\documentclass[twocolumn]{revtex4-1}
\usepackage{amsmath,amssymb,graphicx,bbm}
\usepackage{hyperref}

\begin{document}

\title{Exact time evolution of space- and time-dependent correlation functions after an interaction quench in the 1D Bose gas}

\author{Jorn Mossel and Jean-S\'ebastien Caux}
\affiliation{Institute for Theoretical Physics, University of Amsterdam, Science Park 904, Postbus 94485, 1090 GL Amsterdam, The Netherlands}
\date{\today}

\begin{abstract}
We consider the non-equilibrium dynamics of the interacting Lieb-Liniger gas after instantaneously  switching the interactions off. The subsequent  time evolution of the space- and time-dependent correlation functions is computed exactly.  
Different relaxation behavior is observed for different correlation functions.
The long time average is compared with the predictions of several statistical ensembles. The  generalized Gibbs ensemble  restricted  to a fixed number of particles is shown to give  correct results at large times for all length scales.
\end{abstract}

\maketitle
With the recent advances in ultracold atomic gases it is now possible to realize isolated quantum systems with long coherence times. These experiments are ideal to study non-equilibrium physics \cite{Greiner_Nature_2002, KinoShita_2006_Nat, Hofferberth_Nature_2007}. 
The outcome of these experiments initiated the ongoing debate concerning the circumstances under which  an isolated quantum system will thermalize, which has led to many theoretical studies on non-equilibrium dynamics of quantum systems. Most of these studies considered a  so-called quantum quench: an isolated system is prepared in the ground state of a Hamiltonian $H(c)$, where $c$ is some controllable parameter, like an interaction strength. At $t_0=0$ the parameter $c$ is instantaneously switched to a different value resulting in unitary time evolution governed by the new Hamiltonian. Considerable progress has been made (for a recent review see \cite{Polkovnikov_2011_RMP} ), however there are still many open questions. Under what conditions can a system equilibrate and what is the relevant mechanism? What is the role played by integrability or its absence? Do different kinds of correlation functions show different relaxation behavior? And what is the importance of the initial state?

The difficulty with answering these questions is that one must rely on case by case studies, 
thus, one is faced with the problem of distinguishing  universal behavior from case-specific results. Moreover, most methods used so far suffer from the fact that they are either  valid for short times or for large ones. What we propose here is to study a specific example where we can study the behavior of correlation functions at  all time scales, using an exact method. We can, therefore, not only  determine  what the equilibrium state is, if there is one, but also  how it is reached. Furthermore, the method we use allows us to consider various types of correlation functions, thereby obtaining a more complete picture for this case.

The model we consider is the Lieb-Liniger model, which has experimentally been realized in various circumstances \cite{Kinoshita_2004,KinoShita_2006_Nat, Hofferberth_Nature_2007, Amerongen_2008}. We bring the model out of equilibrium by
 instantaneously turning off the interactions; this is a special case of an interaction quench. For short times one can think of this as a simulation of experiments  where an ultracold Bose gas is released  from a tight transverse confinement  \cite{Imambekov_2009_PRA,Manz_PRA_2010}.

 Several studies on the non-equilibrium dynamics of the Lieb-Liniger model have appeared before.
 The effect of instantaneously turning on the interactions was investigated in
 \cite{Gritsev_2010_JSTAT} using the Bethe Ansatz in combination with a Monte-Carlo sampling technique, and in \cite{Muth_2010_NJP} using the numerical time-evolving block decimation algorithm. The expansion  starting from a regular array was studied in \cite{Lamacraft_2011_PRA}, where the properties of this specific initial state were exploited using the coordinate Bethe Ansatz. In\cite{Sotiriadis_2012} the possibility of studying quenches in the Lieb-Liniger model using an integrable field theory is discussed.

The outline of this article is as follows.  In section \ref{sec:Setup} we describe the setup and explain the  methods to compute correlation functions after the quench. This is followed by section \ref{sec:Results} where we give the results for the time evolution of the  non-local pair  correlation and the auto-correlation. We also introduce a new type of correlation function that is a measure for the correlation between the pre- and post-quench states, which we will call quench-straddling correlations. In section \ref{sec:GGE} we compare the long time average of the correlation functions with the predictions of various statistical ensembles. First we discuss why the Generalized Gibbs (GGE) ensemble fails in this case. Considering the generalized canonical ensemble (GCE),  that is the GGE but with the total number of particles explicitly fixed, we show that it yields the correct results at all length scales. We end with a discussion and conclusions.

\section{Setup and Method}\label{sec:Setup}
The model we discuss in this paper is the one-dimensional Bose gas which is described by the Lieb-Liniger Hamiltonian \cite{Lieb_1963_PR}
\begin{equation}
H(c) = \int_{0}^{L} dx \left\{ \partial_x \Psi^\dagger(x) \partial_x \Psi(x) + c \Psi^\dagger(x) \Psi^\dagger(x) \Psi(x) \Psi(x) \right\}
\end{equation}
with $L$ the length of the system and $c$ the interaction strength. For simplicity we will work with periodic boundary conditions.  We will only consider the repulsive case $c>0$. Two limiting regimes of the Lieb-Liniger model are: noninteracting Bosons  ($c=0$), and the so-called Tonks-Girardeau regime ($c=\infty$)  \cite{Girardeau_1960} where the bosons effectively behave as free fermions. In order to study the non-equilibrium dynamics, we quantum quench the system \cite{Calabrese_PRL_2006,Calabrese_2007_JSTAT}. We do this by first preparing the system in the ground state for $c>0$ and then instantaneously  turning off the interactions: $H(c)\rightarrow H(0)$ at $t_0=0$.

The equal-time 2-point function $\langle \Psi^\dagger(x) \Psi(0)\rangle$ will not evolve in time for this specific quench. The Fourier-transform of  this 2-point function consists of computing $\langle \Psi^\dagger_k \Psi_q \rangle$. Since the states before and after the quench are translationally invariant, one only needs to compute the diagonal components ($q=k$). However, when $c=0$, the momentum occupation operators $\Psi^\dagger_k \Psi_k^{}$ are constants of motion, hence the 2-point function will not be affected by this quench. For higher-point correlations or dynamical ones this will not be the case.  As an illustration we will consider the measurable non-local pair correlation
\begin{equation}
g_2(x) = \frac{\langle \Psi^\dagger(x) \Psi^\dagger(0) \Psi(x) \Psi(0) \rangle}{\langle \Psi^\dagger(0) \Psi(0)\rangle^2}.
\end{equation}
The non-local pair correlation for finite size systems has been computed  in \cite{Caux_2006_PRA} using exact methods for the ground state. The case of finite temperature was studied in \cite{Deauar_2009_PRA} using perturbative techniques. Without loss of generality, we will fix in the following the density at unity $\langle \Psi^\dagger(0) \Psi(0)\rangle = N/L=1$.

In order to study the time evolution after the interactions are turned off, it is useful to write the  final Hamiltonian as $H(0) = \sum_{k} \omega_k \Psi^\dagger_k \Psi_k^{}$ with dispersion relation $\omega_k = k^2$. The time evolution of the field operators readily follows: $e^{iH(0)t} \Psi^\dagger_k e^{-iH(0)t} = \Psi^\dagger_k e^{i k^2 t}$. So instead of  acting with the time evolution operator on the initial state,  we can act with it on the operators. The upshot of this is that we can avoid  decomposing the initial state in terms of eigenstates of the final Hamiltonian. This  decomposition typically involves a number of states that grows exponentially with the total number of particles.  

Using the Fourier-transform of the field operators $\Psi^\dagger(x) = \frac{1}{\sqrt{L}} \sum_{k} e^{ikx} \Psi_k^\dagger$ the Heisenberg picture allows us to express the time evolution of the dynamical correlation function after the quench as
\begin{multline}\nonumber
g_2(x,t_1,t_2) =\langle \phi| \Psi^\dagger(x,t_2) \Psi^\dagger(0,t_1) \Psi(x,t_2) \Psi(0,t_1) |\phi\rangle \\
= \frac{1}{L^2} \sum_{k_1,k_2,k_3} e^{i f(x,t_2,t_1,\{k_i\}) } \langle \phi | \Psi^\dagger_{k_1} \Psi^\dagger_{k_2} \Psi_{k_3} \Psi_{k_1\!+\!k_{2}\!-\!k_3} |\phi\rangle  \label{eq:g2_tevol}
\end{multline}
\begin{equation}\label{eq:g2_tevol} 
f(x,t_2,t_1; \{k_i\})=k_{13}(x\!-\!2k_{23}t_1\!+\!(k_1\!+\!k_3)(t_2\!-\!t_1))
\end{equation}
where $k_{ij} = k_i \!-\! k_j$  and we used $k_1\!+\!k_2\!=\!k_3\!+\!k_4$ because of translational invariance. 
The initial state $|\phi\rangle$ is the ground state of the fully interacting Lieb-Liniger model ($c>0$).
The quench takes place at $t_0=0$ and we evaluate the correlation function at times $0 \leq t_1 \leq t_2$. In equilibrium (at $t_2=t_1=t_0$), this is the normal ordered 2-point function of current $\Psi^\dagger(x)\Psi(x)$ operators and can efficiently be obtained using the techniques described in \cite{Caux_2006_PRA}. In order to study the time evolution, we need to compute a four-point function of field operators $\Psi^\dagger_k$. This  four-point function of the initial state $|\phi\rangle$ can be computed in the framework of the Algebraic Bethe Ansatz \cite{Korepin1993}  by generalizing the methods of \cite{Caux_2007_JSTAT} for their computation of  two-point functions. We will briefly discuss how the computations are done without going in too much detail. First, we express the correlation function in terms of matrix elements. By inserting a resolution of the identity operator between every  pair of adjacent  field operators, one can write the four-point function  in \eqref{eq:g2_tevol} as
\begin{multline}
 \sum_{n_1,n_2,n_3} \langle \phi| \Psi^\dagger_{k_1} |n_1\rangle\langle n_1| \Psi^\dagger_{k_2}  |n_2\rangle \\
 \langle n_2|  \Psi_{k_3}^{}   |n_3\rangle\langle n_3| \Psi_{k_1\!+\!k_2\!-\!k_3}^{} |\phi\rangle
\end{multline}
in terms of the matrix elements $\langle n | \Psi^\dagger_k | m\rangle$   of the field operator. Here $|\phi\rangle$ is the initial state for $N$ particles. The states $|n_1\rangle$, $|n_2\rangle$ and $|n_3\rangle$ are intermediate states with $N\!-\!1$, $N\!-\!2$ and $N\!-\!1$ particles respectively. In order to compute the matrix element  $\langle n | \Psi^\dagger_k | m\rangle$  one first solves the Bethe equations for the states $|n\rangle$ and $|m\rangle$ which results in a set of so-called rapidities for both states. The matrix elements can then be evaluated by computing the determinant of a matrix whose entries are rational functions of the rapidities of the two eigenstates involved. The explicit expression for the matrix element can be found in \cite{Kojima_1997} as a sum of determinants and in \cite{Caux_2007_JSTAT} (see  eq. 41) in terms of a single determinant. What remains to be performed are the actual summations over $n_1$, $n_2$ and $n_3$. Starting from the initial state $|\phi\rangle$ the Fock space of intermediate states $|n_1\rangle$ is scanned by navigating through choices of sets of quantum numbers. For each individual intermediate state, the Bethe equations are solved, and the matrix element is computed. The search for important states is done via the ABACUS method \cite{Caux_2009}, and is close to optimal. In order to verify the accuracy of our results, we keep track of the sum-rule $\sum_{n_1} |\langle \phi | \Psi^\dagger_k | n_1\rangle|^2 = N/L$. The sum over $n_1$ is truncated after a desired precision is achieved. Next, for every state $|n_1\rangle$, we repeat the process by performing a summation over $n_2$, for which we can also compute a sum-rule. We do not need to perform the summation over $n_3$ explicitly because we can use the hermitian conjugate of the matrix element $\langle \phi| \Psi^{\dagger}_{k_1} \Psi^{\dagger}_{k_2} |n_2 \rangle$, thereby constructing the full four-point function. For a system of $N=20$ particles and large interaction $c=1000$, which is the most difficult case, a number  of intermediate states $|n_2\rangle$ of the order of $10^8$  are needed in order  to saturate the sum-rule at $99.6\%$. Once the four-point function is obtained, dynamical correlation functions \eqref{eq:g2_tevol} can be computed straightforwardly.

\section{Results}\label{sec:Results}
In this section, we study the dynamical correlation function $g_2(x,t_1,t_2)$ \eqref{eq:g2_tevol} after the quench, for  all  times. We present the results for a system of size $L=N=20$ starting from the ground state for various $c>0$. For clarity, we specialize to the cases of the non-local pair correlation $g_2(x,t)=g_2(x,t,t)$ and the auto-correlation $g_2(t_1,t_2)=g_2(0,t_1,t_2)$. We conclude this section by considering what we will call the quench-straddling correlations.

\subsection{Non-local pair correlation}
\subsubsection{Short Times}

\begin{figure}[h]
\includegraphics[width=\columnwidth]{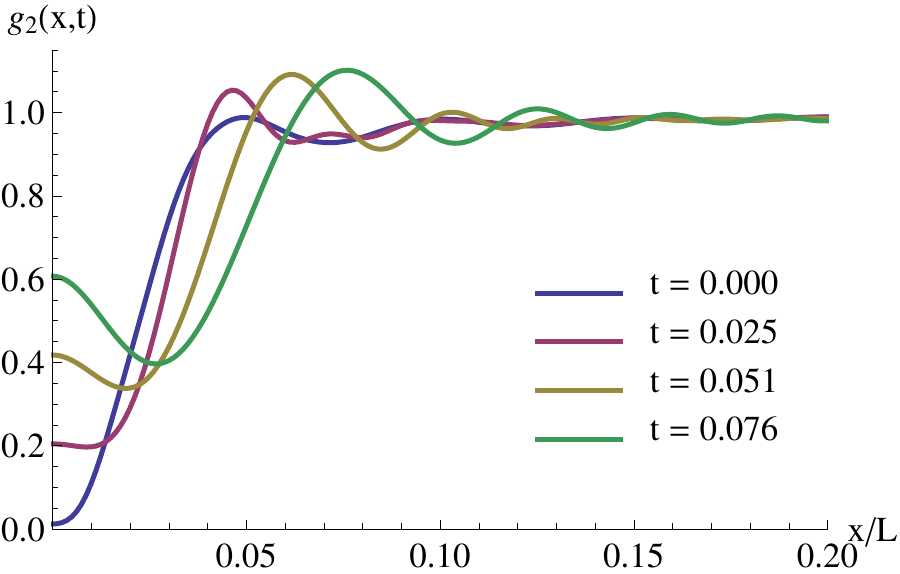}
\caption{$g_2(x,t)$ as a function of $t$  for various $x$. $N,L=20$ and $c=1000$.}
\label{fig:g2x(t)}
\end{figure}

\begin{figure}
\includegraphics[width=\columnwidth]{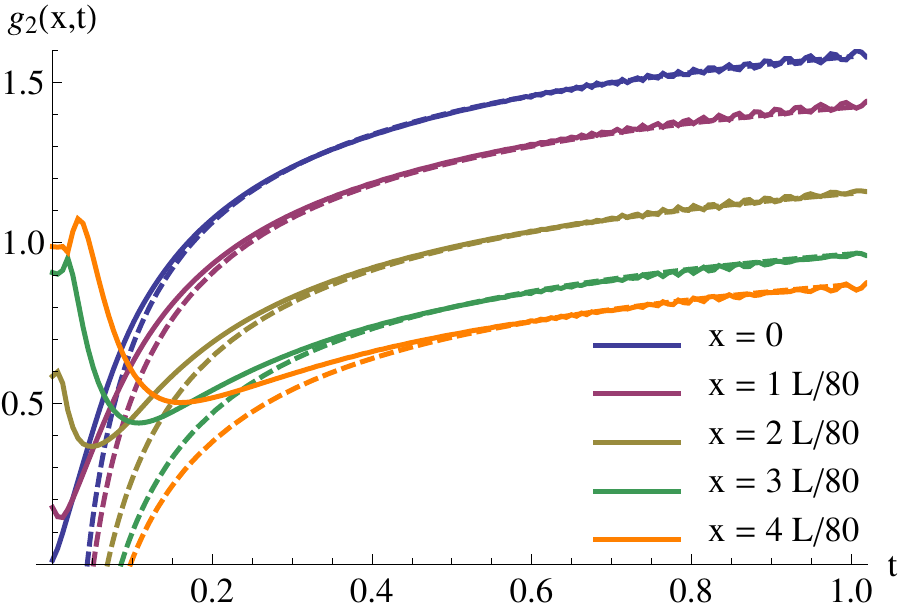}
\caption{$g_2(x,t)$ as function of $x$ (solid lines) for various times together with asymptotic predictions (dashed lines), for $N,L=20$ and $c=1000$.}
\label{fig:g2x(t)fit}
\end{figure}

First, we consider the non-local pair correlation. We plot $g_2(x,t)$  for various times $t$  after the quench as a function of $x$  in fig. \ref{fig:g2x(t)}. For clarity we also plot how $g_2(x,t)$ for  various fixed values of $x$, develops in time in fig. \ref{fig:g2x(t)fit}.
We display here only the results for $c=1000$ because for smaller $c$ the behavior is roughly the same but less pronounced. 
In equilibrium ($t=0$) we see that $g_2(x)$ vanishes for small $x$, as a result of destructive interference due to the fermionic character of the Tonks-Girardeau gas (this is also called anti-bunching). For large $x$, correlations decay,  no destructive interference takes place and $g_2(x) \rightarrow 1$.  If we now focus on the behavior for $x=0$ we see that because of loss of coherence the correlation builds up. Eventually, for large times there will be constructive interference ( $g_2(x)>1$), which is called bunching, and is typical for free bosons  (see fig. \ref{fig:g2x(t)fit}). For small $x$ and small $t$ we also expect that correlations will grow because of loss of coherence, however we have to take into account that $\int_0^L g_2(x,t)dx=1-1/N$ is a conserved quantity which results in non-monotonic behavior for $x>0$ at small times as can be clearly seen in  fig. \ref{fig:g2x(t)fit}.

\begin{figure}
\includegraphics[width=\columnwidth]{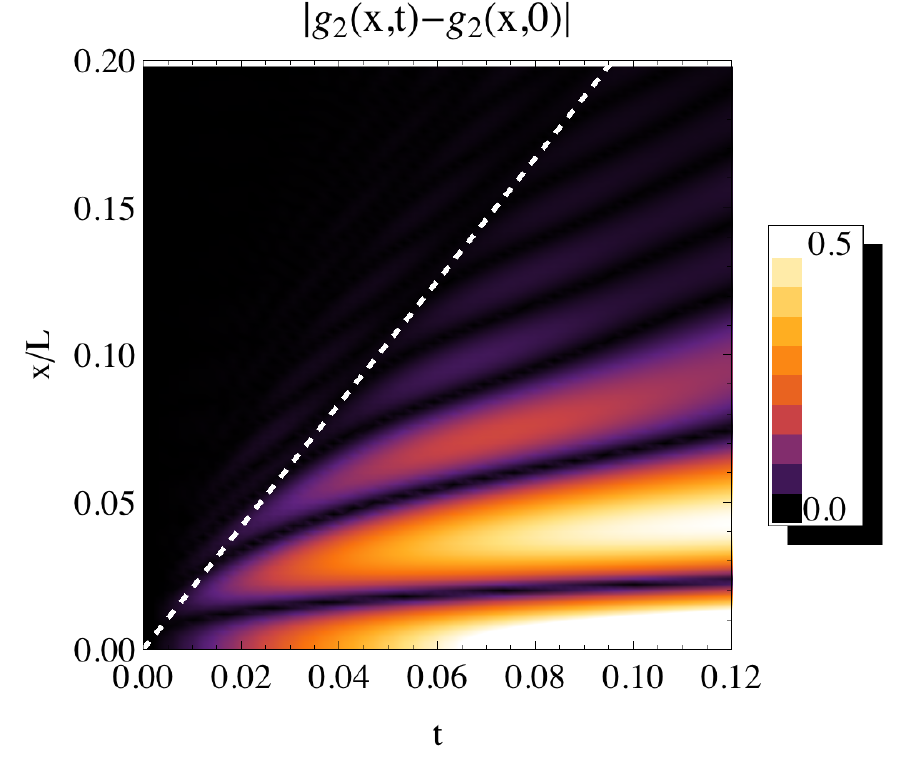}
\caption{$|g_2(x,t)-g_2(x,0)|$, together with $x=v_g t$ (dashed line) for $c=1000$ and $N,L=20$.}
\label{fig:g2_density}
\end{figure}

In order to study how the correlations propagate through the system we plot $|g_2(x,t)-g_2(x,0)|$ in fig. \ref{fig:g2_density}. There is no light-cone effect because of the absence of a maximal velocity.  As reference we have plotted $x=v_{g} t$ (dashed line) where the group velocity is computed as $v_{g} = 2 \sum_{k} |k| \langle \Psi_k^\dagger \Psi_k^{}\rangle$. We see that the dominant changes take place after $t \sim  x/v_g $.

\subsubsection{Finite size effects}
\begin{figure}
\includegraphics[width=\columnwidth]{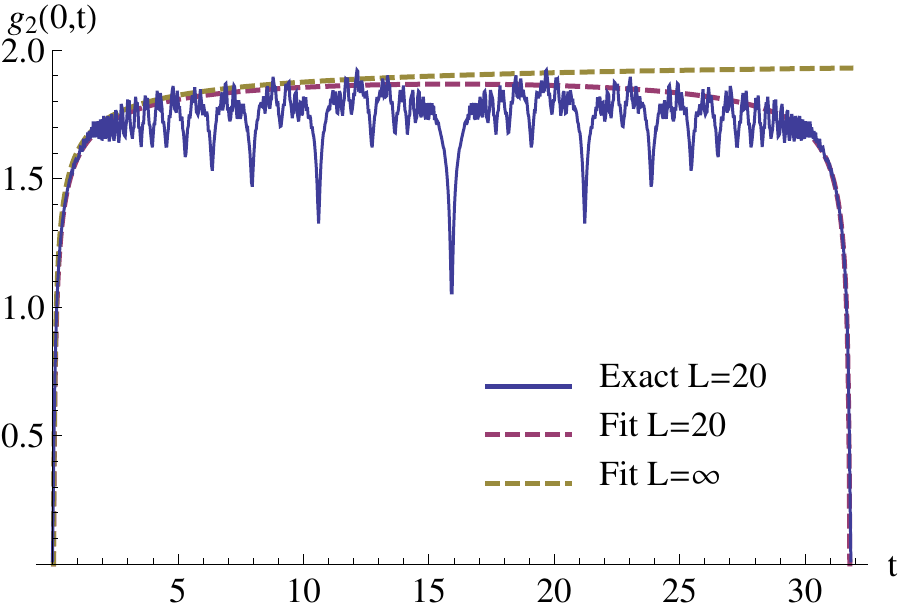}
\caption{Complete and partial revival of $g_2(x=0,t)$ compared with the fits for a system of finite size and infinite size, while keeping the density fixed ($N/L=1$).}
\label{fig:g2N20revival}
\end{figure}

From a finite size study we can conclude that the behavior for short times, discussed in the previous section, is independent of the system size. If we now turn to large times finite size effects become more and more pronounced. Because of the simple dispersion relation $\omega(k_n) = (2\pi n/L )^2$ exact revival occurs at $t_{\text{rev}} = L^2/(4\pi)$.  Partial revivals of decreasing strengths also occur at higher harmonics of the revival time: $t_{\text{rev}}/2, t_{\text{rev}}/3, t_{\text{rev}}/4, \ldots$. Due to the revival, $g_2(0,t)$ gets suppressed for large times, as can be seen in fig. \ref{fig:g2N20revival}. In the thermodynamic limit there will be no suppression and one can easily show that  $g_2(0)=2$ for free bosons in equilibrium.

\subsubsection{Asymptotic behavior}
In order to understand the intermediate time behavior, which is after the non-monotonic behavior  and before finite size effects start to dominate, we fit the results for small $x$ with the following test function
\begin{equation}\label{eq:g2_asymp}
g_2(x,t;c) = a(x;c) - \frac{b(x;c)}{| t_{\text{rev}} \sin (\pi t /t_{\text{rev}}) |^{\alpha(c)}}.
\end{equation}
We motivate this function as follows. The first part $a(x;c)$ is the stationary part. For the time-dependent part we expect algebraic decay, since the initial state is quantum critical. The exact revival, due to the finite size, is taken into account by means of the sine-function. 
The fits for small $x$ and $t$ are compared with the exact result in fig. \ref{fig:g2x(t)fit}. The data suggests that $a(x,c)$ is a decreasing function of $x$ which agrees with the result for the long time average which we will discuss in section \ref{sec:GGE}. We find that $\alpha(c)$  is a decreasing function of $c$ and $\alpha \sim 1/2$ for $c=1000$.  However, since we have only a limited range where we can fit the data and $\alpha$ is small, no firm predictions can be made. After comparing fits for various system sizes, our data seem to be consistent with the dependence on $L$ entering only via $t_{rev}$. This agrees with the intuition that for small $x$ and small times $t$ after the quench finite size effects are irrelevant. Assuming this is indeed the case, one can from the finite size data make predictions for the thermodynamic limit by sending $t_{rev} \rightarrow \infty$ in \eqref{eq:g2_asymp} while keeping the other parameters fixed. 
The stationary part $a(x;c)$ would then be the large time limit in the thermodynamic limit, since the time-dependent part now vanishes for large $t$, which is not the case for finite size systems. For free bosons in equilibrium one can easily show that $g_2(0)=2$, which is compatible with what we find from the fitting data:  $a(0,c) \sim 2$. In fig. \ref{fig:g2N20revival} we compare the exact results for $g_2(0,t)$ with the fit \eqref{eq:g2_asymp} both for finite size and our predictions for the thermodynamic limit.

Since the post-quench state is completely described by the correlations on the initial ground state, one can try to predict asymptotic behavior, such as the exponent in \eqref{eq:g2_asymp}, using  low-energy effective theories like bosonization. If we  write \eqref{eq:g2_tevol} as
\begin{multline}
g_2(x,t) = \frac{1}{L} \sum_{k}   \int_0^L  dz   e^{i k z }\\ \langle \Psi^\dagger(x \!-\! 2k t) \Psi^\dagger(z) \Psi(z\! + \! x \! -\! 2kt) \Psi(0) \rangle,
\end{multline}
one can see that because of the integral over $z$ one needs the correlation function at all length scales. 
Hence whether low-energy descriptions can be used or not remains an open question.

\subsection{Auto-correlation}
\begin{figure}[h]
\includegraphics[width=\columnwidth]{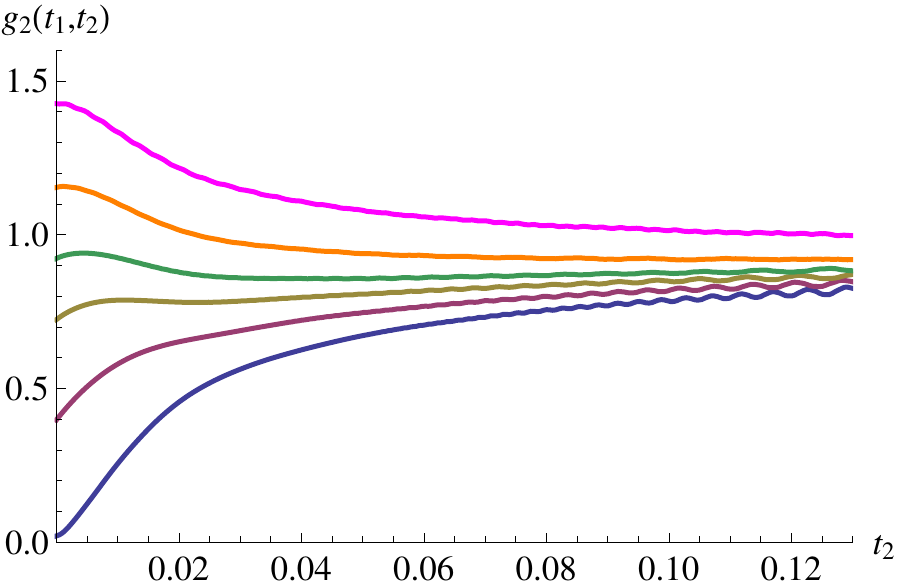}
\caption{The auto-correlation $g_2(0,t_2;t_1)$ for various times $t_1=0.00,0.05,0.10,0.14,0.24,0.53$ as a function of $t_2$  for a system with $N,L=20$  and $c=1000$.}
\label{fig:g2N20autocorr}
\end{figure}

The auto-correlation $g_2(t_1,t_2)$ is plotted in fig. \ref{fig:g2N20autocorr} as a function of $t_2$ after various times $t_1$ after the quench.
As for the non-local pair correlation, there is an anti-bunching bunching transition. However, in contrast to the evolution of the non-local pair correlation, the auto-correlation does increase monotonically in time after the quench, as can be seen in fig. \ref{fig:g2N20autocorr}.
This is consistent with the fact that the integrated auto-correlation is not conserved in time, in contrast with the non-local pair correlations.

\subsection{Quench-straddling correlations}
\begin{figure}
\includegraphics[width=\columnwidth]{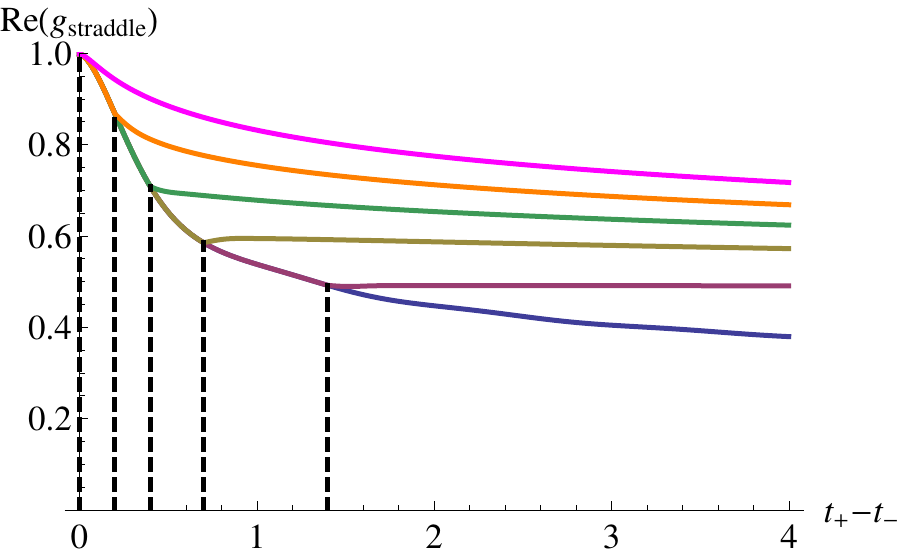}
\caption{The quench-straddling correlation function $\text{Re}\{ g_{\text{straddle}}(0,t_-,t_+)  \}$   as a function of $ t_+-t_-$. The various graphs (from top to bottom) correspond to  the quench times $t_0-t_-=0.0, 0.2, 0.4, 0.7,1.4,\infty$, indicated by the vertical lines. The system parameters are $N,L=80$ and $c=5$. }
\label{fig:g_straddle}
\end{figure}

The versatility of our method allows the computation of various different dynamical correlation functions. As an example we will study a correlation function which is a measure of how the pre- and post-quench states are correlated, what we will call a quench-straddling correlation function.
We consider the situation where the quench takes place at time $t_0=0$. We then compute a two-point function where we evaluate $\Psi(0)$ before the quench   at time $t_{-} < t_0$ and $\Psi^\dagger(x)$ after the quench, at time  $t_{+} > t_0$
\begin{multline}\label{g_straddle}
g_{\text{straddle}}(x,t_-,t_+) = \langle \phi | e^{i H(0) t_+ } \Psi^\dagger(x) e^{-i H(0) t_+} \\
 e^{-i H(c) t_- } \Psi^{}(0) e^{i H(c) t_-}  |\phi \rangle .
\end{multline}
To simplify this expression we first use the Fourier-transform of the field operators $\Psi^\dagger(x) = \frac{1}{\sqrt{L}} \sum_k e^{ikx} \Psi_k^\dagger$. As before, we can then handle the time-evolution of the post-quench Hamiltonian $H(0)$ using the Heisenberg-picture. For the time evolution of the pre-quench Hamiltonian $H(c)$ we use the Schr\"odinger picture; this is achieved by  inserting a resolution of the  identity $\sum_n |n\rangle\langle n|$, in terms of eigenstates $|n\rangle$  of the pre-quench Hamiltonian between $ e^{-i H(0) t_+}$ and $e^{-i H(c) t_- }$. By pulling out all the different phases and using translation invariance we arrive at
\begin{multline}
g_{\text{straddle}}(x,t_-,t_+) =\\  \frac{1}{L}\sum_{n} e^{i (E_n - E_0) t_- + i k_n^2 t_+ + i k_n x} \left|\langle \phi | \Psi_{k_n}^{\dagger} |n\rangle \right|^2.
\end{multline}
Here $E_0$ is the energy of the initial state $|\phi\rangle$ and $E_n$, $k_n$ are the energy and momentum of the intermediate state $|n\rangle$ respectively. In fig. \ref{fig:g_straddle} we plot $\text{Re}\{ g_{\text{straddle}}(0,t_-,t_+)  \}$ for various $t_-$ as a function of $t = t_+$. 
A first observation is that in the case of no quench, $t_+=t_0$, the correlation decays much faster than the case after the quench $t_-=t_0$. 
This can be understood from that fact that, for a sufficiently large interaction strength $c$, the dispersion $E(k)$ is linear for small $k$. Therefore, the majority of the states $|n\rangle$ in \eqref{g_straddle} have an energy  $E(k)>k^2$.
 When considering the general quench times $t_0-t_-$, a striking observation we can make is that just after the quench the correlation function suddenly seems completely relaxed. The explanation is that for a quench-straddling correlation dephasing takes place both via the energy and the momenta, which can be considered as  orthogonal directions. This speeds up the relaxation significantly for a short period right after the quench. The only coherence that is left is for very small $k$, resulting in an extremely slow relaxation compared to the case if there were no quench. 

\section{Statistical  ensembles}\label{sec:GGE}
\begin{figure}
\includegraphics[width=\columnwidth]{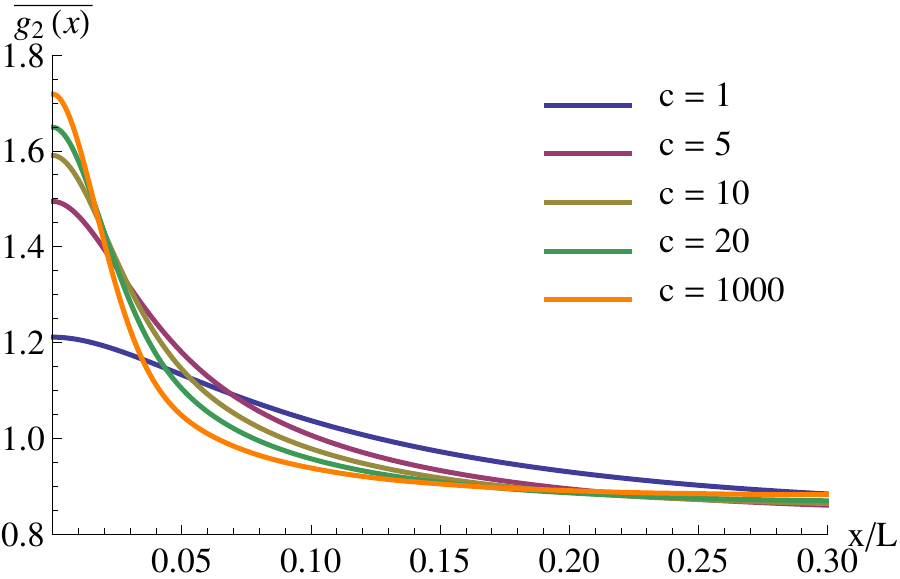}
\caption{The long time average $\overline{g_2(x)}$ for various $c$  for a system with $N,L=20$.}
\label{fig:g2N20LTA(c)}
\end{figure}

In this section we discuss whether the large time behavior of correlation functions after the quench can be described by a statistical ensemble.
For simplicity we restrict ourselves in this section to the analyses of $g_2(x,t)$. Due to the finiteness of the system, no actual relaxation occurs. However, for most of the time, $g_2(x,t)$ oscillates around the same mean value, as is seen in fig. \ref{fig:g2N20revival}. Therefore, it is still useful to consider the long time average (LTA) of correlation functions
\begin{equation}\label{eq:LTA_g2}
\overline{g_2(x;c)} = \lim_{T\rightarrow T_{LTA}} \frac{1}{T} \int_{0}^{T} g_2(x,t;c) dt.
\end{equation}
Note that because of the exact revival we can take  $T_{LTA} =t_{rev}$.
The result of the time average is presented in fig. \ref{fig:g2N20LTA(c)}.
It is well-known that quantum integrable models do not always relax to a state of thermal equilibrium. As an alternative 
 the  Generalized Gibbs ensemble (GGE) \cite{Rigol_PRL_2007} has been proposed. The GGE is based on the idea of Jaynes \cite{Jaynes_1957_PR} to construct a statistical ensemble with maximal entropy subject to constraints for the expectation values of the conserved charges. Expectation values of observables  in the GGE are then computed via
 \begin{equation}
 \langle O(x) \rangle_{GGE} = \text{Tr} \left\{ O(x) e^{-\sum_n \beta_n Q_n} \right\}/Z_{GGE}
 \end{equation}
 where $Z_{GGE}= \text{Tr} \left\{ e^{-\sum_n \beta_n Q_n} \right\}$ is the generalized partition function. The Langrange multipliers $\beta_n$  corresponding to the conserved charges $Q_n$ are fixed via the initial conditions $\langle \phi |Q_n| \phi \rangle = \langle Q_n \rangle_{GGE}$. So far, the GGE has only been used for models which are effectively free, namely the Hamiltonian can be written as $H = \sum_{k} \omega_k \Psi^\dagger_k \Psi_k^{}$,  with  the momentum occupation numbers $\Psi_k^\dagger \Psi_k^{}$ as the obvious choice for the conserved charges. Many successful examples exist  \cite{Cazalilla_PRL_2006,Rigol_PRL_2007, Calabrese_2007_JSTAT, Cramer_2008, Barthel_2008,Kollar_PRA_2008,Fioretto_NJP_2010,Cassidy_2011, Calabrese_PRL_2011}.  Usually, the GGE is implemented as a grand canonical ensemble, which simplifies the results because the partition  function can be written as a product of single states. For example, the partition function can be written as
\begin{equation}
Z_{GGE} = \prod_k (1 \pm e^{-\beta_k} )^{\pm 1}
\end{equation}
where the $+$ and $-$ signs corresponds to fermions and bosons respectively. The Lagrange multipliers are determined via $\langle \phi | \Psi^\dagger_k \Psi^{}_k |\phi\rangle  = 1/(e^{\beta_k}  \pm 1)$, ($+$ fermions and $-$ bosons). However, in this procedure the correlation between different conserved charges are not kept; that is, $\langle \Psi_k^\dagger \Psi_k^{} \Psi_q^\dagger \Psi_q^{} \rangle_{GGE} = 
\langle \Psi_k^\dagger \Psi_k^{} \rangle_{GGE} \langle \Psi_q^\dagger \Psi_q^{} \rangle_{GGE}$ for $k \neq q$.
Despite these shortcomings,  the GGE has been successful in many cases, although exceptions are known, for example when translation invariance is broken \cite{Gangardt_2008_PRA,Caneva_JSTAT_2011}.
Let us now turn to the case of the construction of the GGE for the LTA of $g_2(x)$.   First  we explicitly write down  the LTA of $g_2(x)$ using \eqref{eq:g2_tevol} and \eqref{eq:LTA_g2}
\begin{align}\nonumber
\overline{g_2(x;c)} &= \frac{1}{L^2} \sum_{k_1 \neq k_2}  e^{i x k_{12} }\langle \phi | \Psi_{k_1}^\dagger \Psi_{k_1}^{}  \Psi_{k_2}^\dagger \Psi_{k_2}^{} | \phi \rangle\\
&+ \frac{N(N-1)}{L^2} 
\end{align}
One can see that the LTA for the $g_2(x)$ is explicitly written as a sum over expectation values of products of the conserved charges. From this we can see that the GGE is not applicable in this case, since the initial state is the ground state of an interacting system and Wick's theorem does not apply here. Instead we considered the generalized canonical ensemble (GCE) by keeping the total number of particles fixed. 

\subsection{The generalized canonical ensemble}
Consider a system where the Hamiltonian can be diagonalized in terms of free particles $H = \sum_{k} \omega_k \Psi^\dagger_k \Psi_k^{}$. We impose no restrictions on the dispersion relation $\omega_k$ and the operators $\Psi^\dagger_k$ can have either fermionic or bosonic commutation relations.  The partition function for such a system in the canonical ensemble (CE) can be written as 
\begin{equation}\label{eq:Z_1}
Z_N =\sum_{n_0 = 0}^{n_{max}}  \sum_{n_1 =1}^{n_{max}} \ldots \sum_{n_\infty = 0}^{n_{max}} \prod_{k} e^{- \beta \omega_k n_k} \delta_{\sum_j n_j,N}
\end{equation}
where $n_{max} = 1, N$ for fermions and bosons respectively. The presence of $\delta_{\sum_j n_j,N}$ makes it very complicated to evaluate the sums directly. Fortunately,  one can compute the partition function $Z_N$ for a system of $N$ particles via a well known recursion relation (see for instance  \cite{Landsberg1961,Borrmann_1993_JCP,Wilkens_1997})
\begin{equation}\label{eq:Z_2}
Z_N = \frac{1}{N} \sum_{n=1}^N (\pm1)^{n+1} z_n \; Z_{N-n}
\end{equation}
with $Z_n = 0$ for $n<0$, $Z_0 =1$ and $z_n = \sum_k \exp[ -n \beta \omega_k]$. The minus and plus signs correspond to fermions and bosons respectively. For a given $\beta$ and $N$ the partition function can be evaluated numerically. The computational complexity is of the order of $n_{\text{cutoff}} N^2$, where $n_{\text{cutoff}}$ is the number of one-particle states considered in the computation of $z_n$.
One can easily generalize the computation of the canonical ensemble to what we will call the generalized canonical ensemble (GCE) by making the replacement $\beta \omega_k  \rightarrow \beta_k^{} $. 
Since the partition function now does not factorize, as is the case for the GGE, the expectation values of  products of conserved charges are not necessarily uncorrelated. 
From the partition function one can derive the probability $P^{\geq}_{k}(n)$ of having at least $n$ particles in the state $k$. This is done by starting the summation of $n_k$ at $n$ in \eqref{eq:Z_1} and then using the recursion relation \eqref{eq:Z_2} to obtain
\begin{equation}
P^{\geq}_{k}(n) = e^{-n \beta_k} Z_{N-n}/Z_{N}.
\end{equation}
The probability of having exactly $n$ particles in the state $k$ is therefore
\begin{equation}
p_{k}(n) = \frac{1}{Z_N} \left(  e^{-n \beta_k} Z_{N-n} - e^{-(n+1) \beta_k} Z_{N-n-1} \right).
\end{equation}
Similarly, one can derive the probability of having at least $n_k$ states in $k$ and $n_q$ in $q$ with $k \neq q$:
\begin{equation}
P^{\geq}_{k,q}(n_k,n_q) =e^{-n_k \beta_k-n_q \beta_q} Z_{N-(n_k+n_q)}/Z_N ,
\end{equation}
from which an expression for $p_{k,q}(n_k,n_q)$ can be obtained. Expectation values are now computed as follows:
\begin{align}
\langle \Psi^\dagger_k \Psi_k^{} \rangle_{GCE} &= \sum_{n=1}^N n\; p_k(n).
\end{align}
One can numerically solve these coupled equations in order to obtain values for $\beta_k$. Once all $\beta_k$ are determined we can compute the other expectation values, for instance
\begin{equation}
\langle \Psi_k^\dagger \Psi_k^{} \Psi_q^\dagger \Psi_q^{}  \rangle_{GCE} = \sum_{n_1=1}^{N} \sum_{n_2=1}^{N-n_1} n_1 n_2 \; p_{k,q} (n_1,n_2).
\end{equation}

\subsection{Results}
\begin{figure}
\includegraphics[width=\columnwidth]{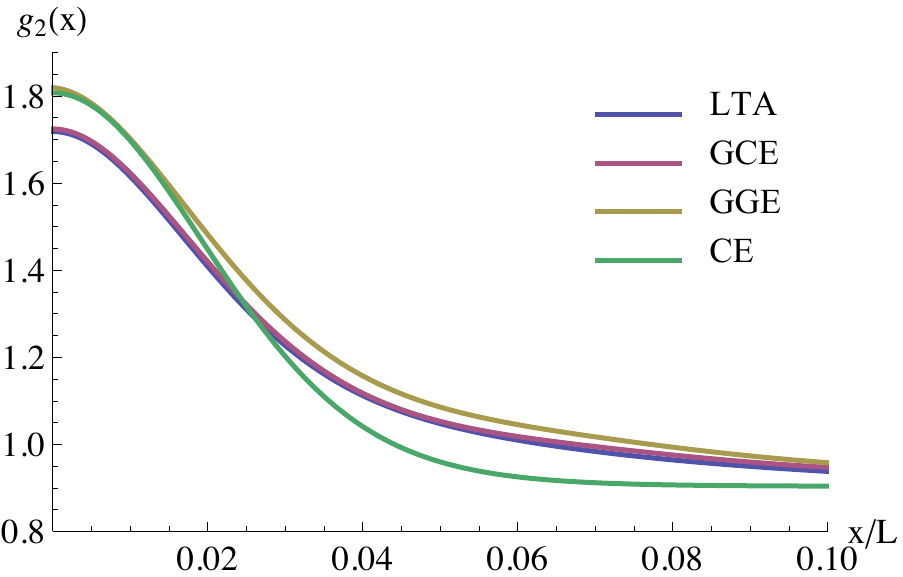}
\caption{The long time average (LTA) of the $g_2(x)$ compared with the predictions of the generalized canonical ensemble (GCE), the generalized Gibbs ensemble (GGE) and the canonical ensemble (CE), for a system with $N,L=20$  and $c=1000$.}
\label{fig:g2x_LTA}
\end{figure}
\begin{figure}
\includegraphics[width=\columnwidth]{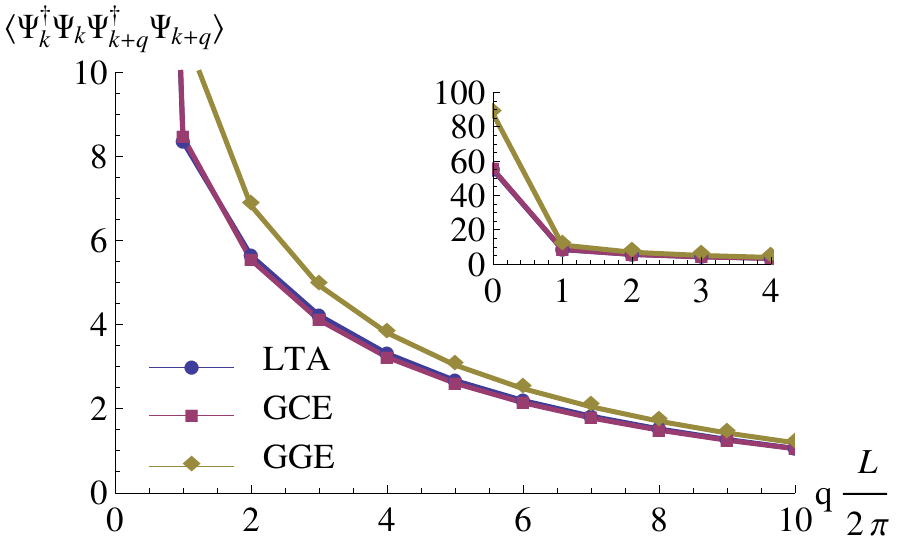}
\includegraphics[width=\columnwidth]{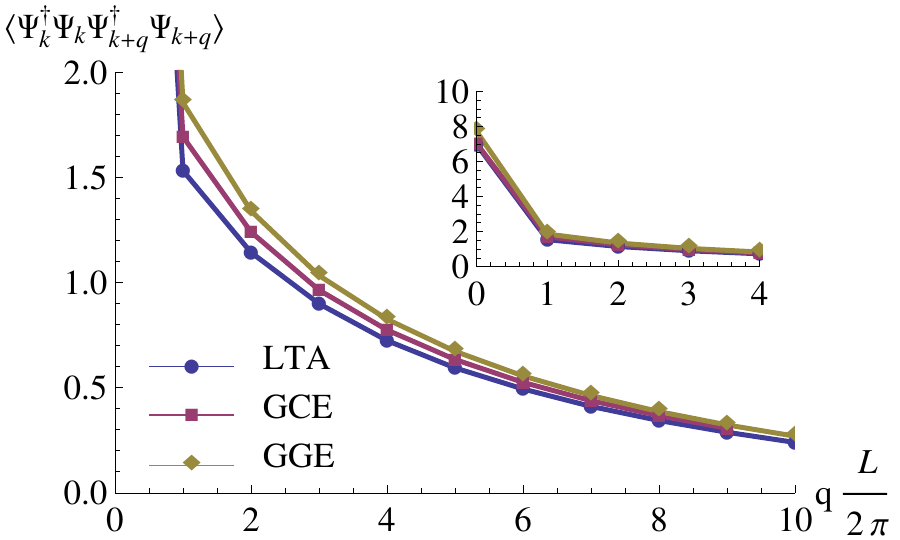}
\caption{Comparison of $\langle \Psi^\dagger_k \Psi_k^{ } \Psi^\dagger_{k\!+\!q} \Psi_{k\!+\!q}^{ } \rangle$ for $k=0$ (top) and $k=2\pi/L$ (bottom) as a function of $q$ for the LTA, GCE and GGE. $N,L=20$ and $c=1000$.}
\label{fig:corr_conserved_charges}
\end{figure}

In fig. \ref{fig:g2x_LTA} we compare the LTA, GGE, GCE and CE for $g_2(x)$. As expected, the predictions from CE and GGE are completely off. The  predictions of GCE   agree extremely well with the LTA see (fig. \ref{fig:g2N20LTA(c)}),
the relative error being less than $0.5\%$ for all $x$. This may be ascribed to the  fact that the sum-rule for $g_2(x)$ is only saturated to $99.6\%$. The reason why the GCE works is not completely obvious. We would like to stress that both for the GGE and the GCE only the expectation values of $\Psi^\dagger_k\Psi_k^{}$ are fixed via corresponding Lagrange multipliers $\beta_k$. In principle one could expand the GGE or GCE by including products of conserved charges  $\Psi_{k_1}^\dagger \Psi_{k_1}^{} \Psi_{k_2}^\dagger \Psi_{k_2}^{} $ with their own Lagrange multiplier $\beta_{k_1,k_2}$,  but that is not what is done here.
Apparently, by demanding that $\langle N^2 \rangle = \langle N \rangle^2$ one almost completely fixes higher order expectation values.  
To gain more insight, we compare the expectation values $\langle \Psi_{k}^\dagger \Psi_{k}^{}  \Psi_{k+q}^\dagger \Psi_{k+q}^{}  \rangle$ for the LTA, GCE and the GGE plotted in fig. \ref{fig:corr_conserved_charges} for $k=0,2\pi/L$. If we first focus on the top graph ($k=0$), we see that for all $q$ the LTA and GCE agree extremely well. The GGE is only valid for large $q>10$, from which we conclude that at this point the conserved charges become uncorrelated. Furthermore, from the inset we see that the fluctuations of $\Psi_0^\dagger \Psi_0$ differ significantly for the GGE. In the bottom graph ($k=2\pi/L$), the correlations are much weaker compared to the case $k=0$. We see that the LTA and GCE differ slightly now, but the GCE is still better than the GGE.

\begin{figure}
\includegraphics[width=\columnwidth]{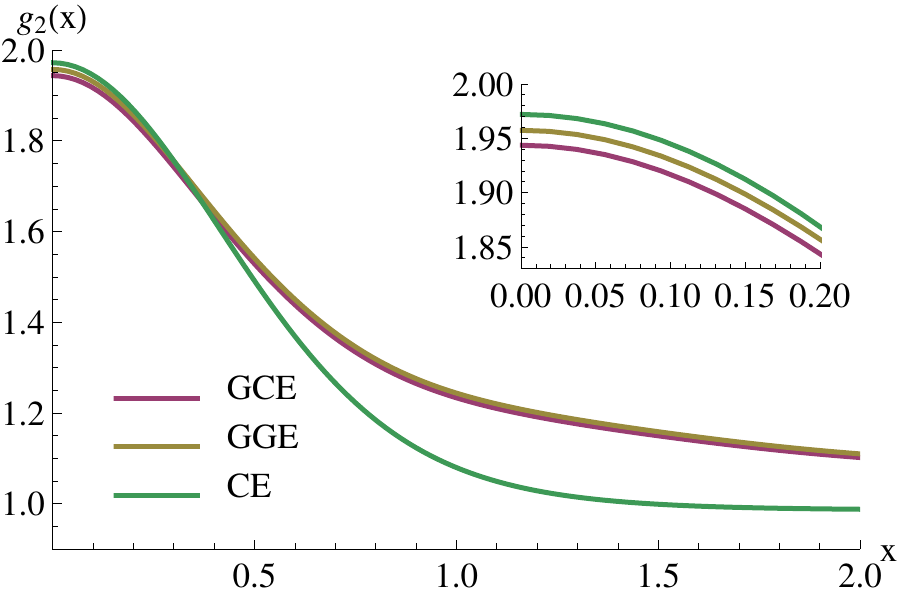}
\caption{The predictions of the GCE, GGE and CE for $N,L=150$  and $c=1000$.}
\label{fig:g2x_GCE_N150}
\end{figure}

The results here are presented for finite size. In the thermodynamic limit one might expect that the prediction of the GCE and GGE coincide for local observables.   On the other hand, correlations like $\langle \Psi_0^\dagger \Psi_0 \Psi^\dagger_q \Psi_q \rangle$ for small $q$ cannot be approximated using Wick's theorem; for these correlations the GGE remains invalid. To study how correlation functions like $g_2(x)$ behave in the thermodynamic limit, we compared the predictions of GCE and GGE for a system with $N,L=150$ in fig. \ref{fig:g2x_GCE_N150}. We see that the difference between GCE and GGE is still present for small $x$, but it is considerably smaller than for the case of $N,L=20$. This leads us to believe  that in the thermodynamic limit the GCE and GGE yield equivalent predictions for $g_2(x)$.

\section{Conclusions and Discussion}
In this paper we have considered an interaction quench by turning off the interactions in the Lieb-Liniger model. By using the Heisenberg picture explicitly, the time evolution was computed via the correlation functions on the fully interacting initial state. The exact method we have used allowed us to study various correlation functions after the quench at all time and length scales. Although this paper focused on the results for the Lieb-Liniger model,  we would like to stress that the methods presented in this paper can be applied to any model for which the Heisenberg picture can be used to efficiently obtain  the time evolution and for which the correlation functions on the initial state can be computed exactly. For example, one could prepare the system in the ground state of the anisotropic Heisenberg spin chain. By  switching off the anisotropy term, the time evolution is accessible from that of free fermions.

In case of the Lieb-Liniger model, we have studied several types of correlation functions after the quench.  As initial state, the ground state of the Lieb-Liniger model was used for various interaction strengths $c>0$. The time behavior of the correlation functions was studied both in the short and long time regimes. As expected, the results are most pronounced for large $c$, although the qualitative features do not strongly depend on the initial interaction strength. As a byproduct we have computed the four-point function of the Lieb-Liniger model for the first time.

The long time average has been compared with various statistical ensembles. As expected, the canonical ensemble fails to make correct predictions, as is usually the case for integrable models. The GGE gives better results. However, for an intermediate number of particles it still fails to capture all the features of the LTA. 
This discrepancy can be understood by the fact that the fluctuation of the total number of particles is large, while for the quench under consideration the total number of particles remains constant. By introducing the GCE, which keeps the total number of particles explicitly fixed, correct predictions for the LTA are obtained. The failure of the GGE can also be explained by the fact that correlations between the conserved charges are not kept. The GCE, which is essentially the GGE with one additional constraint, seems to give the correct correlations between the conserved charges; why this is the case remains unclear on a more formal level.
It is worth to mention that the GCE can be applied in the same cases as the GGE. A firmer test of the GCE would be to study even higher order correlation functions such as $g_n(x) = \langle  (\Psi^\dagger(x))^n (\Psi^{}(0))^n \rangle$. In the thermodynamic limit it is expected that the particle number fluctuations of the GGE become irrelevant for local observables such as the $g_2(x)$; a comparison of GGE and GCE for a large number of particles confirm this.

It would also be interesting to see what the effect is of considering a different basis of conserved charges. For example in \cite{Kollar_PRA_2008} the validity of the GGE was studied using two different bases for the conserved charges,  of which only one  agreed with the long time average. To make a connection with the interacting Lieb-Liniger model with $c>0$, an appropriate basis would consist of the conserved charges in terms of derivatives of the transfer matrix, while sending $c\rightarrow 0$. This leads to the following set of conserved charges: $Q_n = \sum_{k} k^n \Psi_k^\dagger \Psi_k^{}$.  However, for this quench problem the expectation values $\langle \phi | Q_n |\phi \rangle$  for $n \geq 3$ are not well-defined because of the $1/k^4$ behavior of $\langle  \phi | \Psi_k^{\dagger} \Psi^{}_k  | \phi \rangle$ for large $k$ \cite{Olshanii2003} which reflects the ultra-locality of the Lieb-Liniger model. This does not imply that the GGE fails in this case, as one could regularize the results by introducing a lattice spacing $a$ while keeping the system integrable. One can then write down the expectation values as function of $a$ and send $a \rightarrow 0$ at the very end of the calculation. This will be investigated in future work. 

A way to generalize the results of this paper  is by considering quenches starting from arbitrary interaction strength $c_1$ and ending in a different arbitrary $c_2$.
In this case, one cannot use the Heisenberg picture to compute the time evolution of observables as was done in this paper. One could try to solve the time evolution in the Schr\"odinger picture by making a spectral decomposition of the initial state in terms of the eigenstates of the Hamiltonian after the quench. The first problem in this approach is the need for the overlap coefficients between the initial state and final states. For the Lieb-Liniger model, these overlap coefficients are only known in one particular case \cite{Gritsev_2010_JSTAT}; the general problem is still  unsolved. Secondly, the spectral decomposition typically involves an exponential number of states as function of the system size. Unless  there are  huge degeneracies  present in the spectrum, as is only the case for the special points $c=0$ and $c=\infty$, performing the spectral sum seems intractable. We will return to these and further issues in future publications.

\section{Acknowledgements}
This work is part of the research programme of the Foundation for Fundamental Research on Matter (FOM), which is part of the Netherlands Organisation for Scientific Research (NWO).

\bibliographystyle{apsrev}
\bibliography{LL_quench_NJP_2012}

\newpage
\appendix

\end{document}